\documentclass{article}

\usepackage{arxiv}

\usepackage[utf8]{inputenc} 
\usepackage[T1]{fontenc}    
\usepackage{hyperref}       
\usepackage{url}            
\usepackage{booktabs}       
\usepackage{amsmath}
\usepackage{amsfonts}       
\usepackage{nicefrac}       
\usepackage{microtype}      
\usepackage{lipsum}		
\usepackage{graphicx}
\usepackage{natbib}
\usepackage{doi}
\usepackage{authblk}
\usepackage{bm}

\renewcommand{\vec}[1]{\mathbf{#1}}
\newcommand{\prob}{\mathbb{P}}
\newcommand{\PR}{\text{PR}}
\newcommand{\LR}{\text{LR}}
\newcommand{\mat}[1]{\bm{#1}}
\newcommand{\RV}[1]{\bm{#1}}

\renewcommand{\headeright}{}
\renewcommand{\undertitle}{}
\renewcommand{\shorttitle}{Moderators and mediators in conditional attribution}

\date{September 2026}

\begin{document}

\title{On the role of moderator and mediator variables in conditional event attribution of heatwaves over Europe }

\author[1,2,\thanks{Corresponding author: \texttt{s.buschow@fz-juelich.de}}]{Sebastian Buschow} 
\author[1]{Petra Friederichs}
\author[1]{Andreas Hense}

\affil[1]{Institute of Geosciences, University of Bonn, Auf dem Hügel 20, 53121 Bonn, Germany}
\affil[2]{Juelich Supercomputing Centre, Forschungszentrum Juelich, Juelich, Germany}



\maketitle

\begin{abstract}
This study examines how a risk-based attribution analysis of European heat wave events under climate change varies  when explanatory variables are included in the analysis. Instead of relying solely on temperature statistics, the analysis is conducted while conditioning on large-scale circulation or pre-existing climate anomalies.
Using the hot European summers of 2006 and 2007 as examples, we observe several systematic effects. Controlling for the presence of atmospheric blocking explains much of the natural temperature variability and greatly improves the separation of present and pre-industrial distributions in both years. In 2006, the predominant strong blocking raised the expected temperatures, so that, by comparison, the observed anomaly no longer appeared unusually hot, even in a pre-industrial climate. 
More complex effects occur when the conditions are part of the causal chain of climate change, for example, acting as moderators or mediators. By conditioning on such variables, we may remove part of the climate change signal and thus fundamentally alter the outcome of the attribution study. 
Using a simplified model we demonstrate why attribution studies with different kinds of conditions can reach different, seemingly contradictory conclusions. 
\end{abstract}

\section{Introduction}\label{sec:intro}

\begin{figure}
    \centering
    \includegraphics[width=\textwidth]{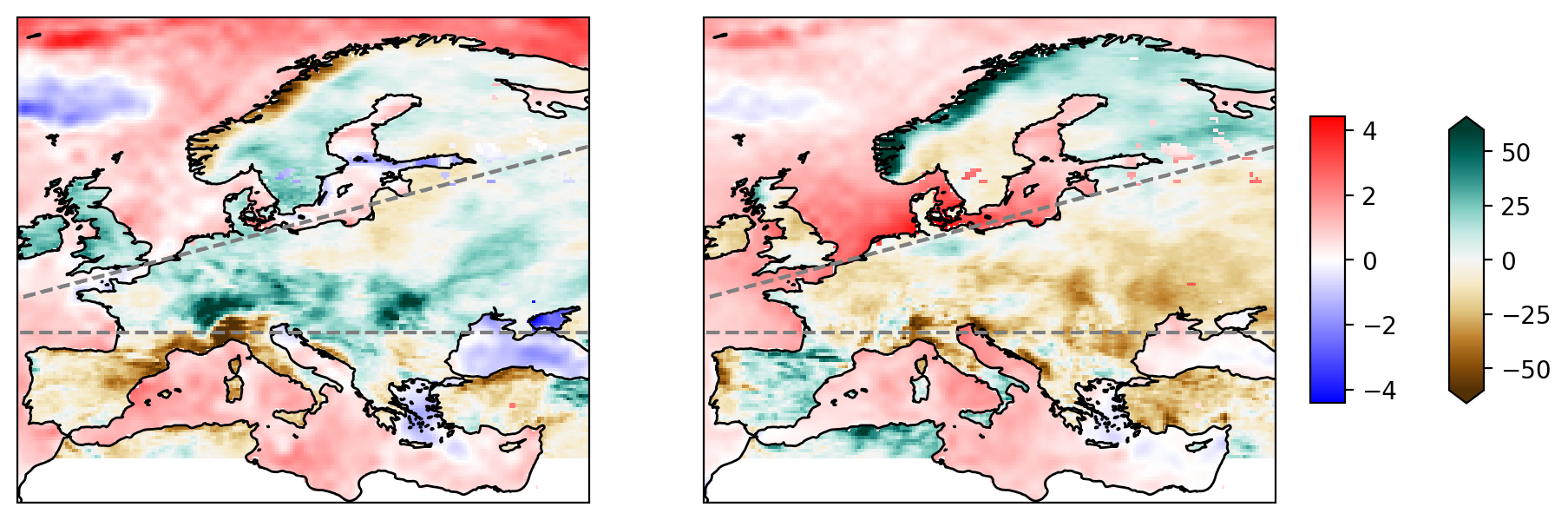}
    \caption{2006 and 2007 northern hemisphere spring (May, April, June) sst and precipitation anomaly in $K$ and $mm/\text{month}$, respectively. Data from ERA5 reanalysis, anomalies with respect to 1961-1990. Dashed grey lines mark the IPCC AR6 region 17 \citep{iturbide2020update} used in our analysis.}
    \label{fig:springs}
\end{figure}

In the aftermath of a heatwave, we often ask whether the event would have unfolded differently in a world without anthropogenic  climate change. Starting with \cite{stott2004human}, event attribution studies have answered this question based on the changing probability distribution of, for example, extreme temperatures in some region of interest. 
In establishing such statistics, very different events will often be  grouped together and treated as samples from one distribution. As an example we consider the summers of 2006 and 2007 in central Europe, both of which were approximately 1.5$^\circ C$ warmer than the climatological mean in the reference period 1961-1990. Figure \ref{fig:springs} shows that the two summers, though with similarly large temperature anomalies, have very different ``backstories'': 2006 saw a  spring with reasonably high precipitation anomalies jointly with only slightly elevated North Atlantic sea surface temperatures (sst). In contrast, the spring of 2007 was exceptionally dry with considerable positive sst anomalies already present in the coastal waters. As we will see below, the two summers also had very different atmospheric circulations, specifically with respect to blocking highs. 
Due to the similar temperature anomalies, attribution studies based on temperature statistics alone will inevitably reach similar conclusions for both years. Nevertheless, it is possible that European summer temperatures with antecedent droughts or strong atmospheric blocking respond very differently to climate change. 

To take such factors into account, we need a \textit{conditional} attribution approach that asks, for example, ``which temperatures would we have observed if the spring and blocking conditions of 2007 had occurred in a pre-industrial climate?'' Such questions are addressed by the so-called \textit{storyline} \citep{trenberth2011attribution, shepherd2016common, shepherd2019storyline} and \textit{analogue} \citep{yiou2007inconsistency} methods of attribution. For a recent overview of different conditional attribution methods, we refer to \cite{thompson2025need}; their common thread is that the properties of an event are studied conditional on some aspects of the climate state that embedded the event in space and time.

The introduction of storylines, in particular, sparked a contentious debate in the scientific literature \citep{lloyd2018climate}, which is also of interest from the perspective of the philosophy of science \citep{lloyd2019climate,garcia2023overstating}. Following this debate, the current best practice is to clearly state which form of conditionality was used, if any, and to combine multiple attribution approaches to gain a more comprehensive understanding of the event. This is advocated by \cite{thompson2025need} and put into practice by \cite{sippel2024could,barnes_climate_2025,leon2025combined, riboldi2026storm,  ermis_contrasting_nodate, clarke_influence_2025} and others. 
These studies present attribution results with different degrees of conditioning and discuss their similarities and differences on a case-by-case basis. 

\begin{figure}
    \centering
    \includegraphics[width=\linewidth]{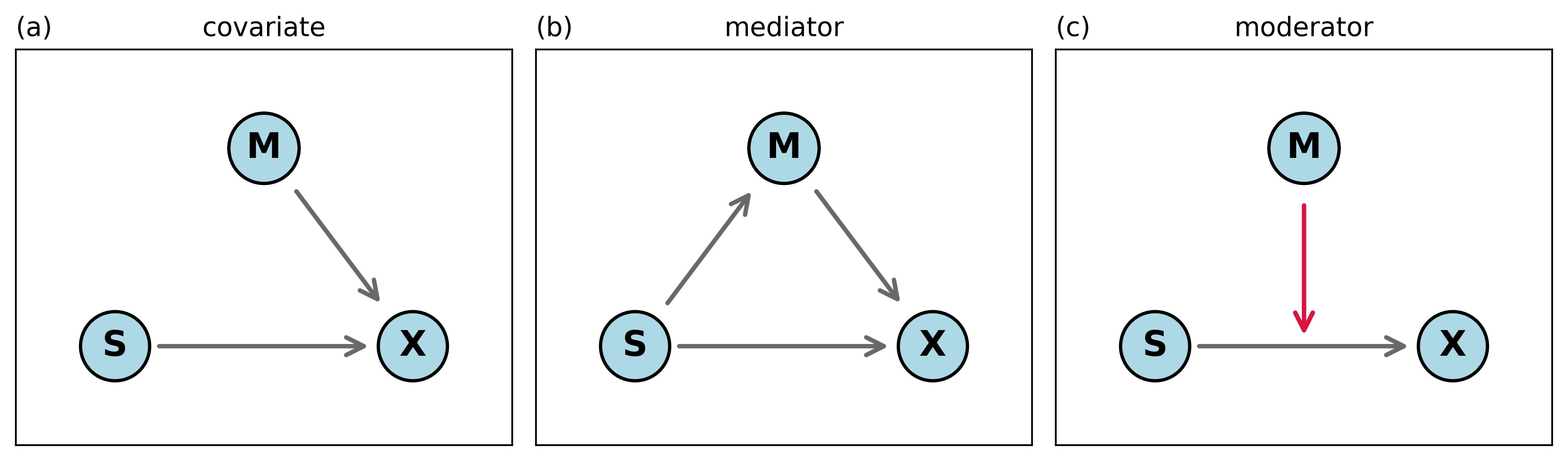}
    \caption{Causal graphs for three archetypal relationships between event variable $X$, climate change scenario $S$ and conditions $M$.}
    \label{fig:modmed_diag}
\end{figure}

However, conditioning can have a significant impact on the outcome of the attribution statement, depending on the specific role a conditioning variable plays within the causal chain in the climate system. Figure \ref{fig:modmed_diag} shows three archetypal roles that a variable can play. First, a variable can function as a covariate (panel a) within the climate system without itself being influenced by climate change or altering the response of the target variable or event to climate change. The second role shown in panel (b) is that of a mediator \citep{hayes_introduction_2013}, through which climate change influences the target variable or event. The conditioning variable represents a process that establishes the causal link between changing greenhouse gas concentrations and the target variable or event. In a third role (c), a variable can act as a moderator that modulates the effect of anthropogenic emissions without representing a necessary condition of the event. 

In this article, these three archetypes serve as  simplified conceptual models to understand the effects of different conditions on a fundamental level. We will address the following research questions:
\begin{enumerate}
    \item What is the impact of conditioning on a variable that accounts for a significant portion of the variability in the event of interest?
    \item How do the results change depending on the causal relationship between event, conditions and forcing? 
\end{enumerate}
As a concrete example, we also ask
\begin{enumerate}
    \item[3.] Did climate change affect the summers of 2006 and 2007 differently, if we take their different stories into account? 
\end{enumerate}

To address these research questions systematically, we use an intermediate approach that incorporates conditioning variables into the traditional statistical event attribution framework (Sect.~\ref{sec:theory}). For the  attribution of the summers 2006 and 2007, we use random effects model (Sect.~\ref{sec:methods}) that allows us to include and combine information from multiple climate models and cover effects of uncertain observations or event definitions. Section \ref{sec:data} summarises the data used in our experiment before we attribute the two summers with and without conditions in Sect.~\ref{sec:res}. The results are discussed in terms of possible causal relationships in Sect.~\ref{sec:diss}, where we consider a  simplified ``toy''-model of mediator and moderator conditions. 
Our main concluding remarks are given in Sect.~\ref{sec:outro}.

\section{Risk-based conditional attribution}\label{sec:theory}

This section and Sect.~\ref{sec:methods} introduce the theoretical underpinnings and practical implementation of risk-based conditional attribution. A list of the main mathematical symbols used is given in Table \ref{tab:symbols} in Appendix \ref{app:symbols}.

Risk-based attribution studies estimate the probability ratio (or risk ratio) for an event $E$ 
\begin{align}
    \PR(E) = \frac{\prob(E|S_1)}{\prob(E|S_0)}\,,\label{eq:PR}
\end{align}
where $S_1$ and $S_0$ are scenarios representing a world with and without climate change, respectively. For example, $E$ could be defined as an observed temperature $X$ exceeding some high threshold $u$. A probability ratio $\PR(X>u)=2$ would then indicate that the risk of temperatures at least as extreme as $u$ has doubled due to climate change. 

We can limit the scope of our attribution study by studying the probability of $E$ conditional on some set of conditions $C$. The outcome is a conditional probability ratio
\begin{align}
    \PR(E|C) = \frac{\prob(E|S_1,C)}{\prob(E|S_0,C)}\,.\label{eq:PRcond}
\end{align}
For instance, if $E$ is again a temperature extreme and $C$ describes the presence of atmospheric blocking, then $\PR(E|C)=3$ would imply that the probability of reaching such extreme temperatures \textit{under blocking conditions} in both scenarios has tripled due to climate change. 

By definition, conditional and unconditional probabilities are linked via $\prob(A|B)=\prob(A,B)/\prob(B)$. Inserting this in Eq.~\ref{eq:PRcond}, we find a simple expression for the conditional probability ratio:
\begin{align}
    \PR(E|C) = \frac{\PR(E,C)}{\PR(C)}\,.\label{eq:PRbayes}
\end{align}
This is a Bayes theorem of probability ratios, which states that the conditional $\PR$ is given by the ratio of the joint $\PR$ of $E$ and $C$ and the $\PR$ of the conditions $C$ alone. 
From this result, we can draw some immediate conclusions:
firstly, if $E$ and $C$ are independent, we can factorise $\PR(E,C)=\PR(E)\PR(C)$, resulting in $\PR(E|C)=\PR(E)$. In the general case, where the conditions affect the event, the conditional $\PR(E|C)$ will differ from $\PR(E)$. When the conditions themselves experience no climate change, we have $\PR(C)=1$ and the conditional $\PR(E|C)$ is equal to the joint $\PR(E,C)$ of event and conditions. Otherwise, the conditional $\PR(E|C)$ is smaller than its joint counterpart when $\PR(C)>1$, i.e., when the conditions themselves become more likely under climate change.

\section{Attribution method}\label{sec:methods}
For a real-world case, both the event $E$ and the conditions $C$ depend on the climate state random vector $\RV{X}$, which could, in general, contain any number of physical variables at various points in space and time. We will describe the observed climate in year $t$ under scenario $S_i$ as
\begin{align}
    \RV{X}_{t,i} = \RV{\mu}_{t,i} + \RV{V}_i + \RV{\epsilon}\,,\label{eq:X}
\end{align}
where $\RV{\mu}_{t,i}$ is the mean state, $\RV{V}_i$ denotes the natural variability and $\RV{\epsilon}$ represents the observational error. Here, we have included the simplifying assumption that $\RV{V}_i$ depends only on the scenario and not on time, while the observation error $\RV{\epsilon}$ depends on neither of those factors.

Since we have only one observation of $\RV{X}_{t,i}$ in the factual world $S_1$, and no observations of the counterfactual world $S_0$, we will estimate the necessary probabilities from ensembles of climate models. The random vector of their modelled climate state for year $t$ under scenario $i$ will be denoted as $\RV{Y}_{t,i,j}$, where the additional index $j$ enumerates the various climate  models.
In the model world, there are no uncertain measurements, but the model mean can differ from $\RV{\mu}_{t,i}$ by a bias $\RV{b}_{t,i}$, i.e.,
\begin{align}
    \RV{Y}_{t,i,j} = \RV{\mu}_{t,i} + \RV{b}_{t,i} + \RV{V}_{i,j}. \label{eq:randomeff}
\end{align}
When multiple model ensembles are considered, each model  $j$ is assumed to have its own realisation of the bias, which could depend on time and scenario as well. Each ensemble member in turn contains an independent realisation of the natural variability $\RV{V}_{i,j}$ for that model and scenario. Equation (\ref{eq:randomeff}) then corresponds to a random effects model (see for example \cite{strochundzwiers} chapter 9.2.8). 

In general, $\RV{X}$ and $\RV{Y}$ could contain the full state of the climate in space and time. To make the problem tractable and easier to interpret, we consider four-dimensional climate states
\begin{align}
    \RV{X}_t = (\text{tas}_t, \text{bl}_t, \text{pr}_t, \text{sst}_t)^T\,,\label{eq:state}
\end{align}
 where the components represent northern-hemispheric summer temperatures, summer blocking, spring precipitation and spring sst for a particular year. Exact definitions are given in Sect.~\ref{sec:data}.

\subsection{Event definition}\label{sub:event}

We can formally relate an event $E$ to the observed climate state $\RV{X}$ via
\begin{align}
    \prob(E|S_i) = \int_\Omega \prob(E|\vec{x})~f(\vec{x}|S_i)~d\vec{x}\,,\label{eq:evdef}
\end{align}
where $\Omega$ is the set of all possible states.
The first factor in the integral contains the event definition, which is the same under both scenarios. The second term is the probability density function (pdf) of the observations we would make under scenario $S_i$. 

Univariate attribution studies often define the event as a threshold exceedance (i.e., as a Heaviside step function)
\begin{align}
    \prob(E|x) = \begin{cases} 1 &\text{if} \hspace{5pt} x>u \\ 0 &\textit{otherwise}  \end{cases}\,.
\end{align}
Inserted into a one-dimensional version of Eq.~(\ref{eq:evdef}), this simply results in
\begin{align}
    \prob(E|S_i) = \int_u^\infty f(x|S_i)~dx = 1-F(u|S_i)\,,
\end{align}
where $F$ is the cumulative distribution function (cdf) of $X$. 

To compute conditional probability ratios in Eq.~(\ref{eq:PRbayes}), we need the joint probability of event and conditions, as well as the probability for the conditions alone. Defining these in terms of component-wise threshold exceedances is not generally appropriate: in our example case in Eq.~\ref{eq:state}, the second component of $\RV{X}$ represents a blocking frequency. Defining $C$ as a threshold exceedance would mean conditioning on all situations \textit{at least as strongly blocked as during the event}, which is not what we want. To avoid this complication, we formally replace the probability in the event definition by a Dirac delta function at the observed value $\vec{x}_o$:
\begin{align}
    \prob(E|\vec{x}) \to \delta(\vec{x} - \vec{x}_o)\,.
\end{align}
Equation (\ref{eq:evdef}) then simplifies to $f(\vec{x}_o|S_i)$ and the probability ratio becomes a ratio of (multivariate) probability densities. To avoid confusion and abuse of notation, we will refer to this as quantity as the \textit{likelihood ratio}
\begin{align}
    \LR(\vec{x}_o) = \frac{f(\vec{x}_o|S_1)}{f(\vec{x}_o|S_0)}\,.\label{eq:dens_ratio}
\end{align}
Throughout this manuscript, we will summarise our attribution results in terms of $\LR$, which is the natural extension of $\PR$ to events that are not threshold exceedance. 


\subsection{Gaussian random effects model}\label{sub:mix}

To evaluate $f(\vec{x})$, we describe the natural variability $\RV{V}_{i}$ in Eq.~(\ref{eq:X}) as a zero mean Gaussian random variable with covariance matrix $\mat{\Sigma}_{V,i}$. For the mean for each year and scenario, we insert an estimator $\hat{\RV{\mu}}_{t,i}$ from our climate models. This estimator is also a Gaussian random variable with error covariance matrix $\mat{\Sigma}_{\hat{\mu},t,i}$ depending on the year and scenario. Lastly, the observation errors $\epsilon$ are assumed to be zero-mean Gaussian with covariance $\mat{\Sigma}_\epsilon$. Assuming that variability, bias and observation error are independent of each other, the desired density is then multivariate normal 
\begin{align}
    f(\vec{x}|S_i) = \mathcal{N}( \vec{x}|\hat{\RV{\mu}}_{t,i}, \mat{\Sigma}_{V,i}+\mat{\Sigma}_{\hat{\mu},t,i}+\mat{\Sigma}_{\epsilon})\,. \label{eq:fx}
\end{align}

The four unknown parameters in Eq.~\ref{eq:fx} are estimated as follows: for each climate model $j=1,...,J$, year $t$ and scenario $i$, we obtain the ensemble means $\bar{\RV{\mu}}_{t,i,j}$. For each model, the natural variability covariance matrix $\mat{\Sigma}_j$ is estimated from a long pre-industrial control run. From these quantities, we estimate $\hat{\RV{\mu}}_{t,i}$ and $\mat{\Sigma}_{\hat{\mu},t,i}$  via maximum likelihood using the \texttt{mixmeta} software package \citep{sera2019extended}.
Following \cite{gasparrini2012multivariate}, the log-likelihood corresponding to the model in Eq.~(\ref{eq:randomeff}) reads
\begin{align}
    \ell(\hat{\RV{\mu}}_{t,i}, \mat{\Sigma}_{b,t,i}) = -\frac{1}{2} \sum_{j=1}^{J}\log(2\pi) +\det(\mat{\Sigma}_j+\mat{\Sigma}_{b,t,i})+(\bar{\RV{\mu}}_{t,i,j}-\hat{\RV{\mu}}_{t,i})^T(\mat{\Sigma}_j+\mat{\Sigma}_{b,t,i})^{-1}(\bar{\RV{\mu}}_{t,i,j}-\hat{\RV{\mu}}_{t,i})\,,\label{eq:loglik}
\end{align}
where $\mat{\Sigma}_{b,t,i}$ is the unknown covariance matrix of the model bias $\RV{b}_{t,i}$. Due to the small number of climate models used (see Table \ref{tab:model_data}), we limit $\mat{\Sigma}_{b,t,i}$ to have a diagonal structure, neglecting  the correlations between biases in different variables.
Equation (\ref{eq:loglik}) is then iteratively optimised with respect to $\hat{\RV{\mu}}_{t,i}$ and $\mat{\Sigma}_{b,t,i}$. The error covariance of $\hat{\RV{\mu}}_{t,i}$ is given by 
\begin{align}
    \mat{\Sigma}_{\hat{\mu},t,i} = \left( \sum_{j=1}^J(\mat{\Sigma}_j+\mat{\Sigma}_{b,t,i})^{-1}\right)^{-1}\,.\label{eq:sigmamufull}
\end{align}

To evaluate Eq.~(\ref{eq:fx}), we also need an estimate of the observed natural variability covariance matrix $\mat{\Sigma}_{V,i}$, which is not a parameter of the random effects model. We can define a natural estimator as the constant covariance matrix $\hat{\mat{\Sigma}}_{V,i}$ by which all $\mat{\Sigma}_j$ could be replaced such that Eq.~(\ref{eq:sigmamufull}) is fulfilled. Thus inserting  $\hat{\mat{\Sigma}}_{V,i}$ into Eq.~(\ref{eq:sigmamufull}), we have 
\begin{align}
    \mat{\Sigma}_{\hat{\mu},t,i} = \frac{\hat{\mat{\Sigma}}_{V,i}+\mat{\Sigma}_{b,t,i}}{J}\,,\label{eq:sigmamu}
\end{align}
which can be re-arranged to obtain
\begin{align}
    \hat{\mat{\Sigma}}_{V,i} := J\cdot\mat{\Sigma}_{\hat{\mu},t,i} - \mat{\Sigma}_{b,t,i}\,.\label{eq:sigmaV}
\end{align}
This is inserted as our best estimate of $\mat{\Sigma}_{V,i}$ in Eq.~(\ref{eq:fx}). 
To check that the covariance matrix defined in this way is positive semi-definite (and thus valid), we have to verify that
\begin{align}
x^T ~\hat{\mat{\Sigma}}_{V,i} ~x \geq 0 
~~\Leftrightarrow~~ x^T (J\mat{\Sigma}_{\hat{\mu},t,i}) x  \geq x^T ~\mat{\Sigma}_{b,t,i} ~x\,.\label{eq:posdef} 
\end{align}
From Eq.~(\ref{eq:sigmamufull}), we recognise that $J\mat{\Sigma}_{\hat{\mu},t,i}$ is the harmonic mean over matrices $(\mat{\Sigma}_j+\mat{\Sigma}_{b,t,i})_{ j=1,...,J}$. Since each $\mat{\Sigma}_j$ is positive semi-definite by definition, Eq.~(\ref{eq:posdef}) just expresses that fact that the harmonic mean (like any other mean) is monotone increasing in the sense of the Loewner order: adding some positive semi-definite  $\mat{\Sigma}_j$ to each element will not decrease the mean \citep{Bhatia_chapter4_2009}. 

We note that
\begin{enumerate}
    \item If the bias term is negligible, Eq.~(\ref{eq:sigmaV}) simplifies to the arithmetic mean of the individual covariance matrices $\mat{\Sigma}_j$. 
    \item Due to $\mat{\Sigma}_{b,t,i}$, the resulting covariance matrix of $\RV{X}$ can differ between scenarios and years, even though we assume a single, fixed $\mat{\Sigma}_{j}$ per model.
    \item The number of ensemble members is intentionally not used in estimating the overall mean and covariance. We thereby avoid giving extra credence to models from centres with more computational resources (``ensemble of opportunity'' problem, see \citet{tebaldi2007use}). In particular, an unequal number of members per scenario (see Table \ref{tab:model_data}) would artificially modify the probability ratios. 
\end{enumerate}

Lastly, we need an estimate of the observation error matrix $\mat{\Sigma}_\epsilon$. We obtain a rough estimate of this quantity from the discrepancies between four global reanalysis datasets (see Sect.~\ref{sec:data}). Because of this small sample size, we cannot rely on the naive estimates for any individual year. Instead, we compute the six unique time series of differences (as a pre-whitening operation) between pairs of datasets and estimate $\mat{\Sigma}_\epsilon$ from the pooled sample of these differences (analogous to the ensemble dressing approach in \cite{scholzel2011probabilistic}). 

With all terms in Eq.~(\ref{eq:fx}) estimated, we can evaluate the individual and  joint likelihood ratio $\LR$ for our variables of interest. According to Eq.~(\ref{eq:PRbayes}), the conditional $\LR$s can then simply be computed as the ratio between joint and unconditional $\LR$s.

\subsection{Bootstrap confidence intervals}\label{sub:boot}

To contextualise our estimates of conditional and unconditional $\LR$, we need to assess their uncertainty resulting from the randomly realised model biases and member states. To this end, we  perform a parametric bootstrap based on the model given in Eq.~(\ref{eq:randomeff}): for each year $t$, scenario $i$ and model $j$, we draw a realisation of $\RV{b}_{t,i}$ using the estimated $\mat{\Sigma}_{b,t,i}$. Then, the same number of members as in the original ensemble (see Table \ref{tab:model_data} below) is sampled from  $\mathcal{N}(\cdot|\RV{b}_{t,i},\mat{\Sigma}_j)$. With these resampled ensembles, we repeat the same estimation procedure as before to obtain 100 bootstrap samples of the various $\LR$s.

\section{Data}\label{sec:data}

As mentioned in the introduction, our real-world example revisits the classic case of European summer temperatures, studied in the very first event attribution study \citep{stott2004human}. More specifically, we consider spatial means of daily mean temperature in the IPCC AR6  region 17 (central Europe, see Fig.~\ref{fig:springs}), averaged over the Northern Hemisphere (NH) summer months June to August as our event variable.
To represent the dynamical conditions during each summer, we include the mean frequency of blocking or subtropical ridge activity, averaged over the same region and timespan. Some details about the blocking definition following \cite{sousa2021new} are given in Appendix \ref{app:block}.
Two further variables are added to represent the pre-existing conditions at the beginning of each summer season: first, the central European precipitation total during the preceding NH spring (March to May), and second, the NH spring mean sst averaged over the European region (Fig.~\ref{fig:springs}). For 2006 and 2007, we have presented these spring conditions in Fig.~\ref{fig:springs}. All variables, their regions, units and abbreviations are summarised in Table \ref{tab:variables}.

\begin{table}
    \centering
     \caption{Aggregated variables used in this study. For the regions, as defined in \cite{iturbide2020update}, see also Fig.~\ref{fig:springs}.}
    \label{tab:variables}
    \begin{tabular}{lllll}
         &  full name & spatial aggregation & time aggregation & unit\\\hline
        tas & daily mean temperature & AR6 reg. 17 & JJA mean & $^\circ C$ \\
        bl & presence of block or ridge &  AR6 reg. 17 & JJA sum & days / summer \\
        pr & precipitation & AR6 reg. 17 & MAM mean & $mm$ / month\\
        sst & sea surface temperature & AR6 reg. 16,17,19 &  MAM mean & $^\circ C$\\
    \end{tabular}
\end{table}

Since we are interested in the interplay between these variables, we need a representation of the observed state in which all four are physically consistent. We therefore employ ERA5 reanalysis data \citep{hersbach2020era5} as our best estimate of the observed NH spring and summer conditions. 
While ERA5 does provide a 10-member ensemble used during data assimilation, the variance across members is miniscule in recent years for the large-scale averages we study (not shown). Since further uncertainty may result from the assimilation system and choice of observations, we instead obtain additional data from three global reanalyses: NCEP2 \citep{kanamitsu2002ncep}, MERRA2 \citep{gelaro2017modern} and JRA3Q \citep{kosaka2024jra}. The error covariance matrix is estimated from the common time period  1980-2024, following the approach described in Sect.~\ref{sub:mix}.

\begin{table}
    \centering
    \caption{Number of hist- and hist-nat members and piControl years for the CMIP6 ensembles used in this study.}
    \label{tab:model_data}
    \begin{tabular}{lccc}
        {model} & {hist-nat} & {historical} & {piControl (\# years)}\\
        \hline
        canesm5       & 42               & 35        & 451         \\
        cesm2         & 1                & 10        & 1200         \\
        cnrm-cm6-1    & 3                & 23        & 213         \\
        hadgem3-gc31-ll & 4              & 4         & 500         \\
        ipsl-cm6a-lr  & 6                & 23        & 115         \\
        miroc6        & 3                & 10       & 500          \\
        mri-esm2-0    & 5                & 5        & 200          \\
    \end{tabular}
\end{table}

For the climate state under scenarios $S_0$ and $S_1$, we rely on CMIP6 simulations  of the historic and hist-nat scenario, respectively \citep{eyring2016overview, gillett2016detection}. Our choice of models is mainly limited by the need for daily geopotential height fields for the blocking and ridge algorithm. Table \ref{tab:model_data} lists the included models and the used number of members from each scenario. In addition to these scenario runs, we use the respective piControl simulations to compute the natural variability covariance matrix of each model $\mat{\Sigma}_j$ . 

\begin{figure}
    \centering
    \includegraphics[width=.6\linewidth]{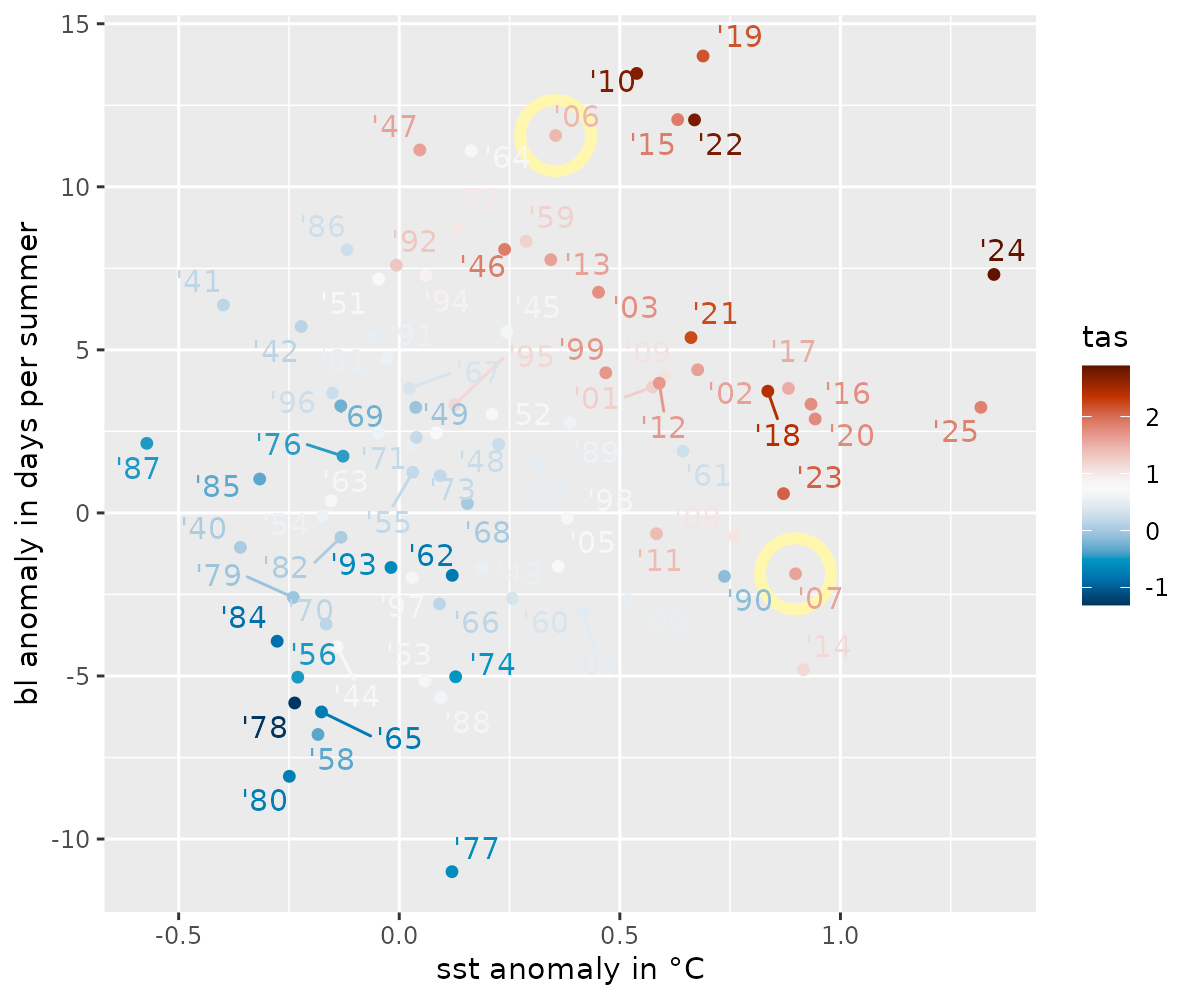}
    \caption{Central European summer temperature (colour), as a function of European sst during the preceding spring (x-axis) and summer mean blocking and ridge activity (y-axis) in ERA5 data. All three variables are anomalies with respect to the reference period 1961-1990.}
    \label{fig:scatter}
\end{figure}

All variables are converted to anomalies by subtracting the average values from the climate reference period 1961-1990 for each model and scenario as well as the reanalyses, separately. To first order, this step effectively removes the model biases in the event and especially in the conditions, and focuses our analysis on the climate change signals alone.

Figure \ref{fig:scatter} displays the ERA5 anomalies of sst, bl and tas. As expected, anomalously hot NH summers in Europe are associated both with strong blocking and positive antecedent sst anomalies. Many of the most recent years show positive anomalies in both tas and sst, indicating an observed climate change signal in both variables. In contrast, a shift towards stronger or weaker blocking is not readily apparent. The two  summers of 2006 and 2007 to be discussed are encircled in Fig.~\ref{fig:scatter}.

\section{Attribution of two European summers}\label{sec:res}

Figure \ref{fig:scatter} shows the strong contrast between our two selected NH summers from Fig.~\ref{fig:springs}: the summer of 2006 (tas anomaly of $+1.4^\circ C$) has one of the strongest blocking anomalies on record (ten days more than average), with relatively low antecedent sst and above average spring precipitation. In contrast, the following summer of 2007 was slightly warmer than 2006, but without anomalous blocking activity. It was preceded by a relatively dry spring with exceptionally high regional sst anomalies. 

For both years the ensembles means and piControl covariance matrices of our models are combined by the random-effects model described in Sect.~\ref{sec:methods}. Before discussing the resulting distributions, 
we can look at the correlation matrices corresponding to our estimates of natural variability $\hat{\mat{\Sigma}}_{V,i}$. For the historical scenario in 2006, we find
\begin{align}
    \mat{C}_ {2006,1} = \begin{array}{r|rrrr}
        & \text{tas} & \text{bl} & \text{pr} & \text{sst} \\
          \text{tas} & 1.00 & 0.76 & -0.21 & 0.22 \\ 
          \text{bl} & 0.76 & 1.00 & -0.08 & 0.00 \\ 
          \text{pr} & -0.21 & -0.08 & 1.00 & -0.07 \\ 
          \text{sst} & 0.22 & 0.00 & -0.07 & 1.00 \\
\end{array}
\end{align}
As expected, NH summer blocking is a strong predictor of temperatures in central Europe, explaining over half of the variance. Precipitation shows a slight anti-correlation with tas, reflecting a weak link between dry NH springs and hot NH summers. The correlation between tas and sst has a slightly larger magnitude and opposite sign, indicating that pre-existing positive sst anomalies in the European oceans favour a warm NH summer in central Europe.
We find almost identical matrices for the year 2007, as well as the hist-nat scenario in both years. While this is not guaranteed to be the case, it is hardly surprising since our computation of anomalies has removed much of the inter-model variance, resulting in near-zero between model covariances.

\begin{figure}
    \centering
    \includegraphics[width=\linewidth]{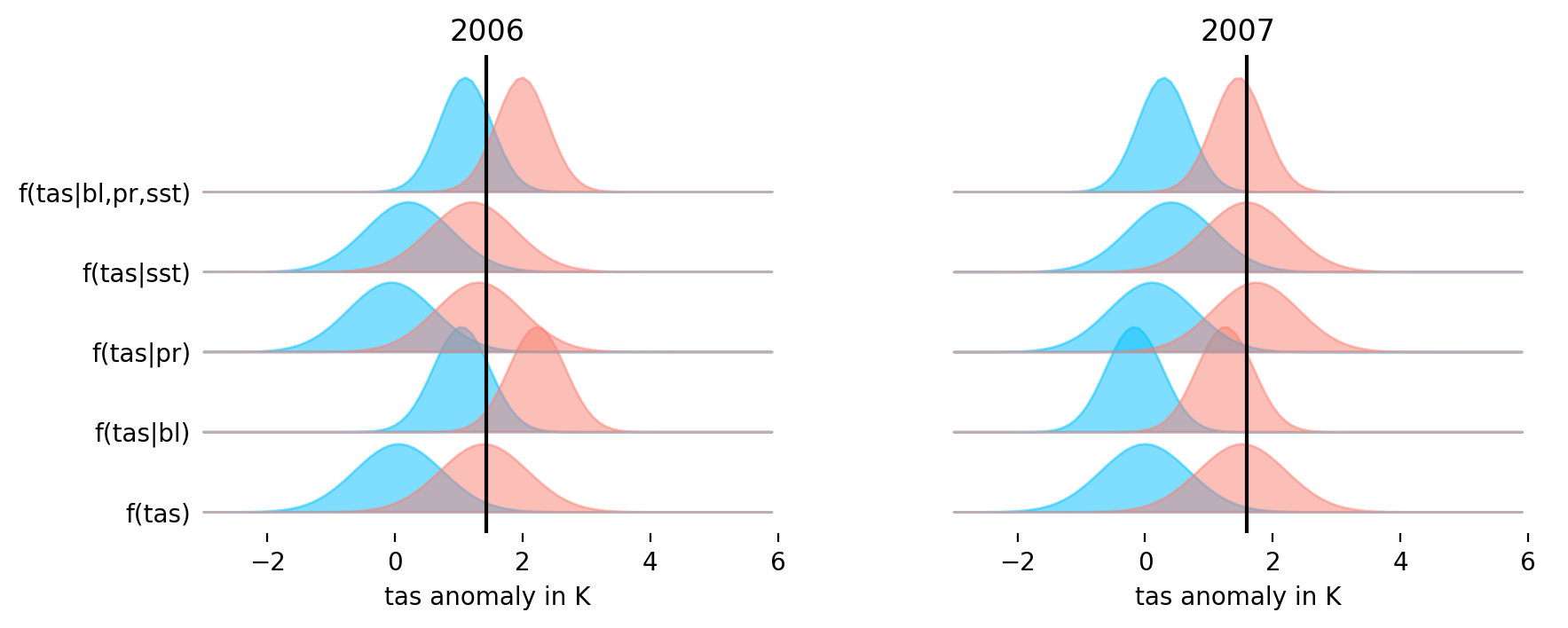}
    \caption{Probability density functions of tas anomalies in 2006 and 2007 under scenario $S_1$ (red) and $S_0$ (blue). From bottom to top: unconditional density, density conditional on each individual condition and conditional on all the conditions variables. Vertical lines mark the ERA5 temperature anomalies in the respective years.}
    \label{fig:ridgelines}
\end{figure}

\begin{figure}
    \centering
    \includegraphics[width=\linewidth]{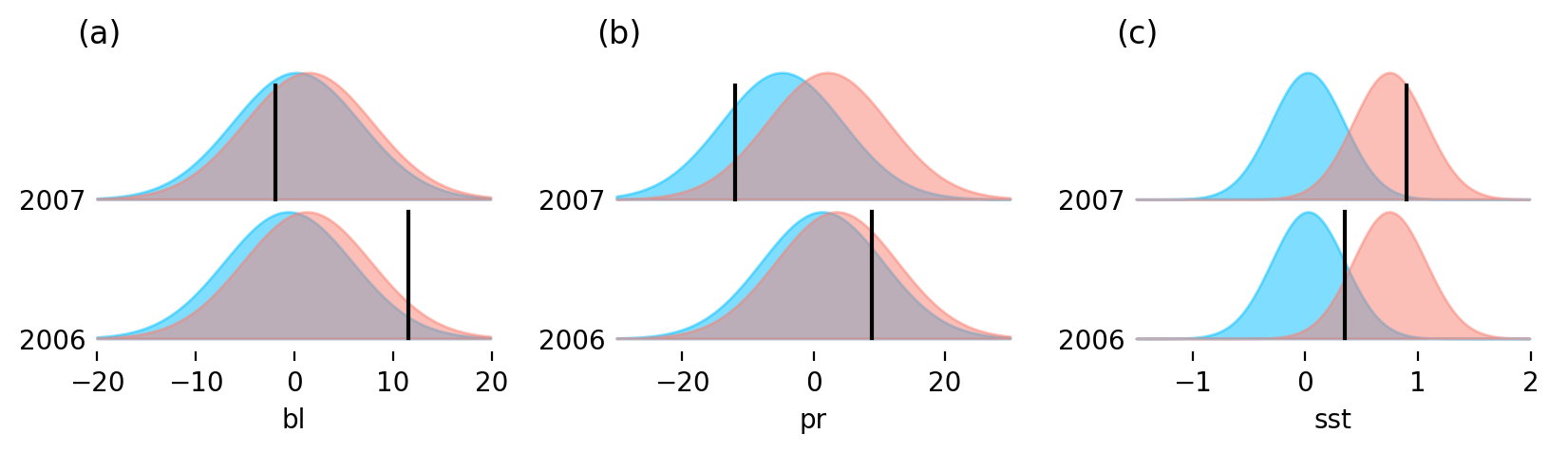}
    \caption{Unconditional pdfs of blocking (a), precipitation (b) and sea surface temperature anomalies (c) in 2006 and 2007 under scenario $S_1$ (red) and $S_0$ (blue). Vertical lines mark the ERA5 values for the respective years and variables.}
    \label{fig:ridgelines_others}
\end{figure}

We now turn our attention to the resulting conditional and unconditional temperature distributions for the two years shown in Fig.~\ref{fig:ridgelines}. To better understand the effects of the three condition variables, Fig.~\ref{fig:ridgelines_others} shows their distributions together with the ERA5 values in the two years. 

For 2006, Fig.~\ref{fig:ridgelines} shows that the observed temperature value is close to the unconditional expectation under the historical scenario (red), but far in the tail of the hist-nat distribution (blue). In terms of the likelihood ratio, $\LR(\text{tas})\gg 1$. 
This result is flipped completely when we condition the distribution on blocking: firstly, $f(\text{tas}|\text{bl})$ is considerably narrower than the unconditional pdf. In addition, the highly unusual blocking activity (Fig.~\ref{fig:ridgelines_others} a) shifts the distribution under both scenarios to higher temperatures. As a result, the 2006 value is actually more typical for hist-nat than for the historical scenario, which implies $\LR(\text{tas}|\text{bl})<1$!

The positive precipitation anomaly (Fig.~\ref{fig:ridgelines_others} b) has hardly any impact on the temperature distribution, but sst shows a significant effect: due to the strong sst warming (Fig.~\ref{fig:ridgelines_others} c), the small observed anomaly is rather unusual for the historical scenario (panel d). Consequently the distribution of tas conditional on sst is shifted to lower values for the hist scenario. If we consider this condition in isolation, it also reduces the likelihood ratio, but not as strongly as blocking. However, when we combine all three conditions (top row of Fig.~\ref{fig:ridgelines}), the effects of strong blocking and lower sst compensate each other. As a result, the 2006 temperature is close to the intersection of the conditional pdfs, i.e.,
\begin{align}
    \LR_{2006}(\text{tas}|\text{bl},\text{pr},\text{sst}) \approx 1\,.
\end{align}

For the NH summer 2007 the attribution result is very different (right part of Fig.~\ref{fig:ridgelines}). Here, conditioning on blocking also reduces the variance of the temperature distribution, but does not lead to strong shift since the observed blocking value is close to the climatological mean (Fig.~\ref{fig:ridgelines_others} a). As a result, the difference between the two pdfs is massively enhanced, resulting in a much stronger conditional attribution statement (Fig.~\ref{fig:PRs} below). 

The effect of precipitation conditions is again hardly visible. While the anomaly has a similar magnitude but opposite sign from 2007, Fig.~\ref{fig:ridgelines_others}\,b appears to indicate a slightly stronger climate change signal in the precipitation distribution itself. However after inspection of the underlying model timeseries, we interpret this as a random fluctuation in the hist-nat simulation. Due to the relatively small sample size for that scenario (Table \ref{tab:model_data}) and the inherent uncertainty of precipitation in coarsely resolved climate simulations,  we will not interpret this phenomenon any further. 

The rather high sst anomaly in 2007 is close to normal for the historical scenario but far too large for hist-nat (Fig.~\ref{fig:ridgelines_others} c). The corresponding conditional temperature distribution $f(\text{tas}|\text{sst})$ is therefore shifted towards its historical counterpart (reduced $\LR$). 
When we condition tas on the ``whole story'' (top right curves in Fig.~\ref{fig:ridgelines}), the effects again compensate slightly. However, the variance reduction is clearly dominant, meaning
\begin{align}
    \LR_{2007}(\text{tas}|\text{bl},\text{pr},\text{sst}) \gg \LR_{2007}(\text{tas})\,.
\end{align}

\begin{figure}
    \centering
    \includegraphics[width=0.6\linewidth]{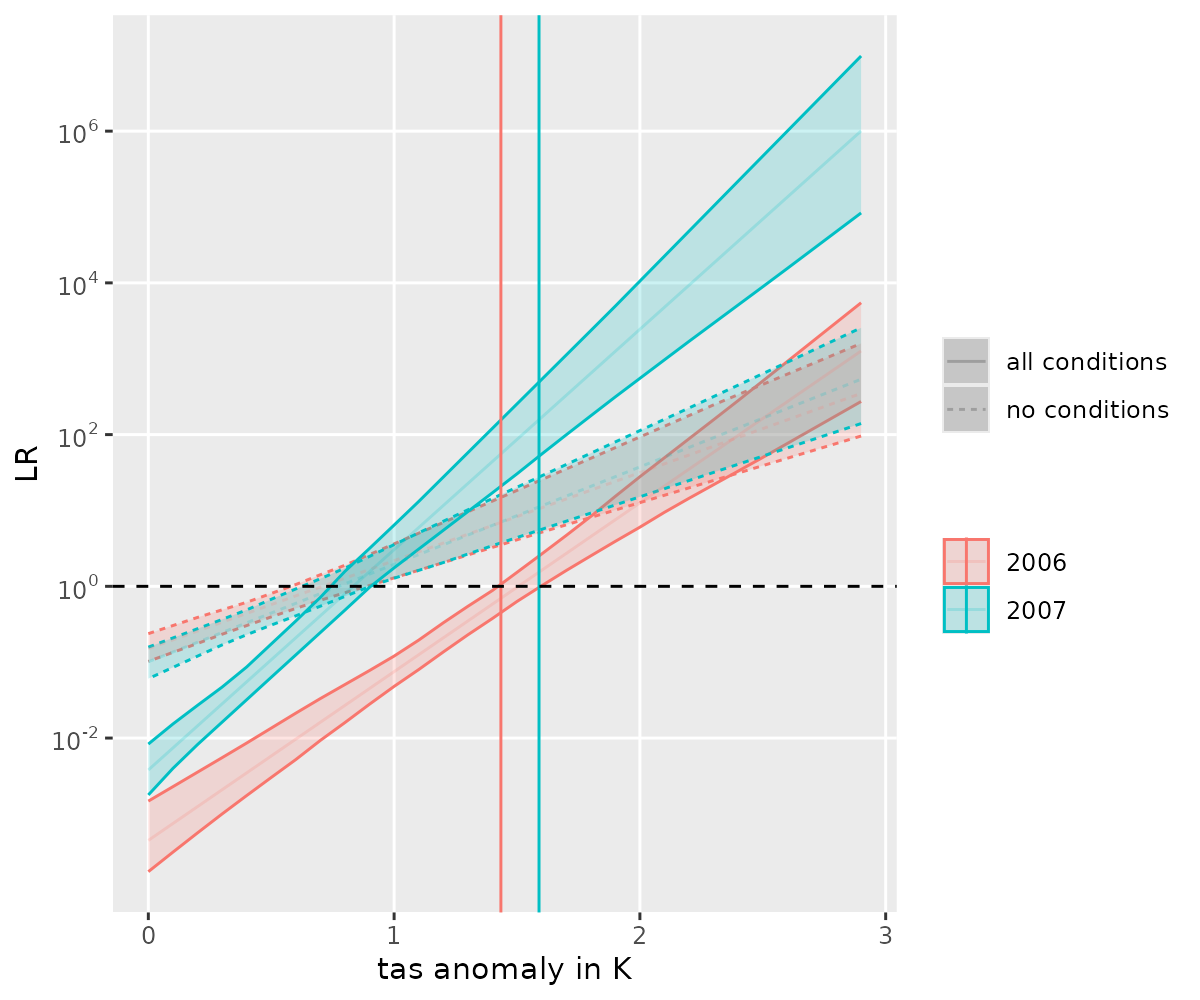}
    \caption{Likelihood ratios as a function of tas for the years 2006 and 2007, given no additional conditions or all three conditions (pr, bl, sst), as in Fig.~\ref{fig:ridgelines}. Vertical lines mark the corresponding ERA5 values for the two years, areas denote 95\,\% confidence intervals based on 100 bootstrap resamples.}
    \label{fig:PRcurves}
\end{figure}

The outcome of our conditional and unconditional attribution experiments for 2006 and 2007 is quantitatively summarised in terms of likelihood ratios in Fig.~\ref{fig:PRcurves}. We observe that all conditional and unconditional $\LR$s are exponential functions of the temperature anomaly (linear on the logarithmic axis). The  unconditional curves are nearly identical for the two years, resulting in $\LR\approx 10$ for the observed values. The curves for the conditional $\LR$ are steeper and have opposite offsets for the two years. As a result, the confidence interval for 2006 includes $\LR=1$, while the conditional $\LR$ lies around 100 in 2007. Hence, when we  control for the dynamics and the preconditions in 2006, there is no evidence for an increased event likelihood. In 2007, conditioning on the same variables leads to an increase in $\LR$ by orders of magnitude. 

\begin{figure}
    \centering
    \includegraphics[width=0.7\linewidth]{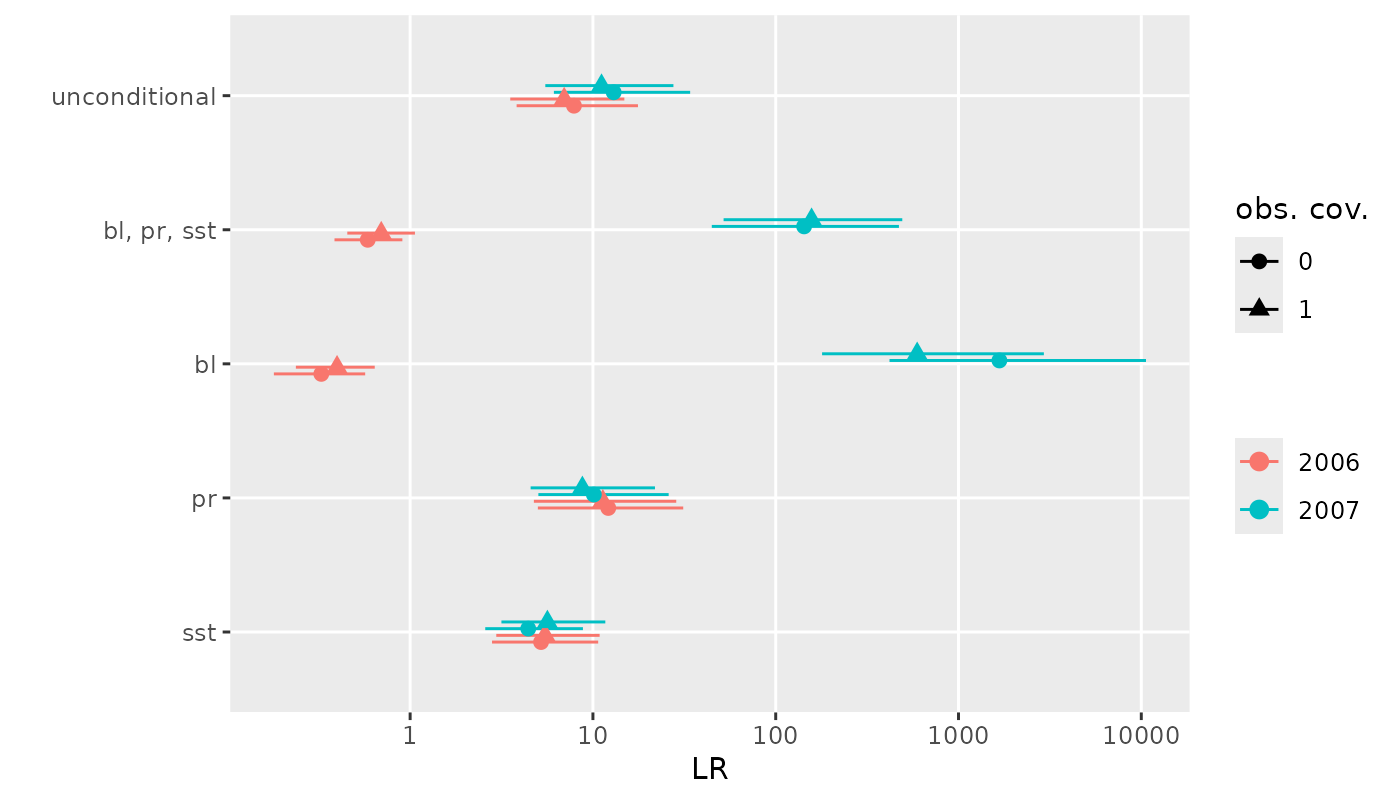}
    \caption{Mean and 95\,\% confidence interval of probability ratios for the NH summer temperatures in 2006 and 2007, estimated from 100 parametric bootstrap resamples. The points in the top row show unconditional $\LR$s, values below are conditional on the variables marked on the y-axis. Circles and triangles correspond to results with and without the observation error covariance $\mat{\Sigma}_\epsilon$.}
    \label{fig:PRs}
\end{figure}

For completeness, Fig.~\ref{fig:PRs} summarises the $\LR$ for the observed values given all or no conditions (as in Fig.~\ref{fig:PRcurves}) and conditional on each variable individually. As expected from the previous discussion, we see a massive effect from the reduced variance when conditioning on bl. This is slightly compensated by the other two conditions, with sst in particular resulting in nearly identical $\LR$ for both years. 

Lastly, Fig.~\ref{fig:PRs} also shows the effect of the observation errors $\mat{\Sigma}_\epsilon$. The difference in attribution outcomes with and without uncertain observations (triangles and circles in Fig.~\ref{fig:PRs}) is very limited in our example. The overall effect, seen most clearly in the extremely uncertain $\LR(\text{tas}|\text{bl})$, tends to bring the ratios closer to unity by ever so slightly reducing the signal to noise ratio. We expect this effect to become much more relevant for smaller spatio-temporal scales and data-sparse regions outside of Europe.

\section{Role of the conditioning variable}\label{sec:diss}

We now want to understand and classify the effects of conditioning seen in Sect.\ref{sec:res} in terms of the different archetypal causal relations, denoted as pure covariate, mediator, and moderator (Fig.~\ref{fig:modmed_diag}).
To this end, we consider a simple model for a univariate target variable $X$ and a univariate conditioning variable $M$ with the following assumptions:
\begin{itemize}
    \item All process variables are normally distributed with variances independent of the scenario.
    \item $M$ has an additive effect on $X$, which may vary depending on the scenario.
    \item Other processes affecting $X$ are represented by a random variable $\varepsilon$ which is independent of $M$.
\end{itemize}
The corresponding model is represented by 
\begin{align}
    X_i = \mu_X + \varepsilon + i\cdot \Delta x + \beta_1 M_i + i\cdot\beta_2M_i\,,\label{eq:modmedmodel}
\end{align}
where $\varepsilon$ is a zero-mean Gaussian random variable, $i\in\{0,1\}$ again denotes the scenario $S_i$, and $\mu_X,\Delta x,\beta_1$,and $\beta_2$ are constant parameters. 
\begin{align}
    M_i = \eta + i\cdot \Delta m\label{eq:modmed_M}
\end{align}
represents our condition variable, where $\eta$ follows zero-mean Gaussian distribution independent of $\epsilon$, and $\Delta m$ is a constant climate change signal in $M$.  For simplicity, we have also assumed that $M$ has mean zero in the counterfactual scenario. It follows, that both the unconditional distribution of $X_i$ and the conditional distribution $X_i|M_i=m$ are also Gaussian, and the results of our investigation are fully determined by the conditional and unconditional means and variances. They are given by
\begin{alignat}{3}
     &E[X_i] &&= \mu_X + i\cdot[ \Delta x + (\beta_1+i\cdot \beta_2)\Delta m ]\label{eq:E} \\
   & E[X_i|M_i=m] &&= \mu_X + \beta_1 m + i\cdot[\Delta x + \beta_2m]\label{eq:E|M_old} \\
    &\textit{Var}[X_i]\label{eq:V} &&= \textit{Var}[\varepsilon] + (\beta_1 + i\beta_2)^2\textit{Var}[\eta]\\
    &\textit{Var}[X_i|M_i=m] &&=\textit{Var}[\varepsilon] \,.\label{eq:V|M}
\end{alignat}

\begin{figure}
    \centering
    \includegraphics[width=.3\linewidth]{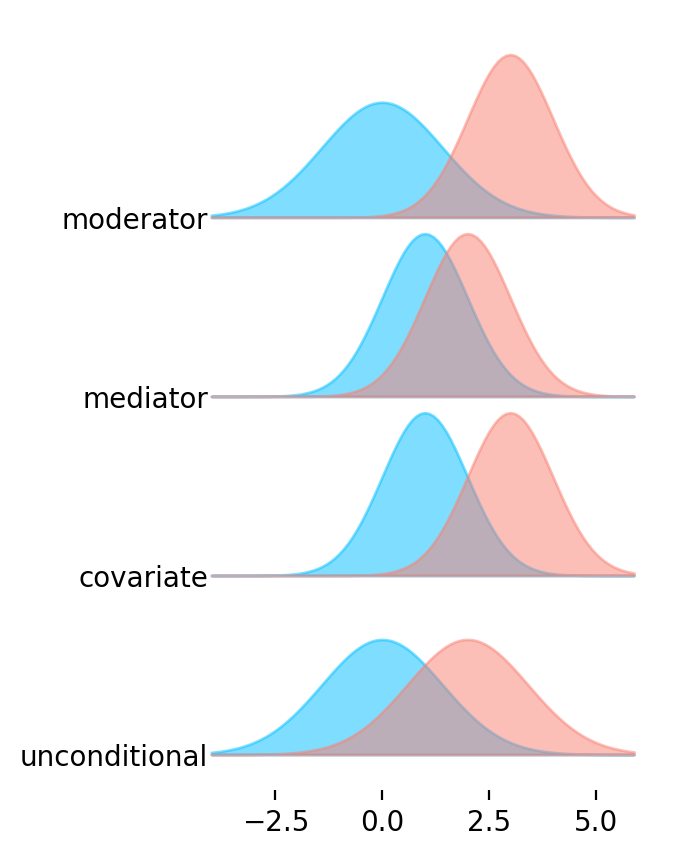}
    \caption{Probability densities corresponding to Eq.~(\ref{eq:modmedmodel}). \textit{Bottom row}: unconditional pdfs for $S_0$ (blue) and $S_1$ (red). \textit{Above}: examples corresponding to conditional pdfs for the three causal situations from Fig.~\ref{fig:modmed_diag} (see text).}
    \label{fig:modmed_pdfs}
\end{figure}

With this simplified model, we now study the three archetypal causal relationships between climate change, event and conditions, shown in Fig.~\ref{fig:modmed_diag}. For these examples, it is convenient to keep the unconditional distributions at the same location in all three cases, thus mirroring the situation in Sect.~\ref{sec:res} (i.e., one unconditional distribution, and multiple possible conditions). This is achieved by re-parametrising the model in terms of the unconditional expectations $E[X_0],E[X_1]$ instead of $\mu_x,\Delta x$. The conditional expectation values then read
\begin{align}
    E[X_i|M_i=m] = E[X_i] +(\beta_1+i\beta_2)m - (\beta_1+\beta_2)i\Delta m\,,\label{eq:E|M}
\end{align}
where we have tacitly inserted $i^2=i$ since $i\in\{0,1\}$.
For our example distributions shown in Fig.~\ref{fig:modmed_pdfs}, we set $E[X_0]=0$, $E[X_1]=2$, $\textit{Var}[\varepsilon]=\textit{Var}[\eta]=1$ and consider a condition value of $m=1$. The parameters $\beta_1,\beta_2,\Delta m$ are chosen to reflect the different causal situations:
\begin{itemize}
    \item For the pure covariate, $\beta_1=1$, $\beta_2=0$, $\Delta m=0$.
    \item For the pure mediator, $\beta_1=1$, $\beta_2=0$, $\Delta m=1$.
    \item For the pure moderator, $\beta_1=0$, $\beta_2=1$, $\Delta m=0$.
\end{itemize}

\subsection{Conditions act as a pure covariate}\label{sec:purecov}
In this setting, the climate change scenario $S_i$ and the conditions $M$ act on $X$ completely independently (Fig.~\ref{fig:modmed_diag}\,a). In our simple model this is realised if there is no climate change signal in the conditions ($\Delta m=0$) and no interaction term between $S_i$ and $M$ (thus $\beta_2=0$).
Consequently, the conditional distributions are shifted by $\beta_1m$ in both scenarios, and the conditioning reduces the variance by $\beta_1^2\textit{Var}[\eta]$. The resulting pdfs are shown in Fig.~\ref{fig:modmed_pdfs}.

This is approximately the situation we observed in Sect.~\ref{sec:res} when conditioning tas on blocking activity, which is largely unaffected by climate change: the distribution under both scenarios becomes narrower and receives a shift. Since in our toy model the variance remains unaffected by climate change, we have a simple equation for the log-likelihood ratio of the conditional probability densities (see Appendix \ref{app:toy}), consisting of three components:
\begin{align}
    \log \LR(x|m) =\underbrace{\left(x-\frac{E[X_0]+E[X_1]}{2} - \beta_1 m\right)}_{(i)} \cdot \underbrace{(E[X_1]-E[X_0])}_{(ii)} \cdot\underbrace{(\textit{Var}[X]-\beta_1^2\textit{Var}[\eta])^{-1}}_{(iii)}\label{eq:logPRcov}
\end{align}
We note that the unconditional $\LR(x)$ is recovered by $\beta_1=0$.
The first term $(i)$ in Eq.~(\ref{eq:logPRcov}) represents the anomaly of $x$ with respect to the conditional mean averaged over the two scenarios. As we have seen in Fig.~\ref{fig:PRcurves}, $\log\LR$ scales linearly with $x$ and is shifted by $m$. The second factor $(ii)$ is the climate change signal of $X$. Factor $(iii)$ describes the reduction of $X$'s variance due to conditioning on $M$, which  is always positive. If we assume a positive shift $\Delta x>0$ due to climate change, then $\LR>1$ holds if and only if the event under investigation $x$ is unusually large even conditioned on $m$. This explains how conditioning on $M$ can modify and  even flip the attribution statement. For example if we control for very favourable conditions $\beta_1m\gg0$, we may find that $x$ was actually unusually small for the climate change scenario $S_1$ and more typical of the counterfactual climate state $S_0$.
Equation (\ref{eq:logPRcov}) also explains the general increase in $\LR$ we observe when conditioning on a strong predictor like blocking: the larger $\beta_1$, the larger is the part of the variance explained by the conditions, leading to a larger $\LR$ due to term $(iii)$. 

\subsection{Conditions act as a pure mediator}
If a variable is acting as a mediator, it is causally located between $S$ and $X$ (see Fig.~\ref{fig:modmed_diag}\,b) and forms part of the causal pathway through which $X$ is changing  \citep{igartua_mediation_2021}. 
In our simplified case, this implies a non-zero climate change signal in $M$ ($\Delta m\neq 0$). If $M$ is purely a mediator, the interaction term is absent ($\beta_2=0$).

The reduction of the variances is then the same as for a pure covariate as defined in Sect.~\ref{sec:purecov}. The changes in conditional and unconditional expectation value (Eq.~(\ref{eq:E|M})) are now related via
\begin{align}
    E[X_1|m]-E[X_0|m] = (E[X_1]-E[X_0]) - \beta_1\Delta m\label{eq:deltaXmed}
\end{align}
Assuming, for simplicity, that $\beta_1\Delta m>0$, i.e., conditions shift in a way that favours large $X$, the conditional climate change signal is smaller than the unconditional one, because we have removed the part of the effect that is mediated by the conditioning variable $M$. 

The likelihood ratio from Eq.~(\ref{eq:logPRcov}) is modified in two opposing ways: in term $(i)$, $m$ is replaced by $m-\Delta m/2$, accounting for the fact that $E[M]$ now depends on the scenario. For $\beta_1\Delta m>0$ this effect would increase $\LR$. Conversely, term $(ii)$ is reduced by $\beta_1\Delta m$, thus reducing $\LR$. For very unusual conditions ($m\gg\Delta m$), the effect in $(i)$ can be neglected and the net effect of conditioning on a mediator will be a reduction of $\LR$. 
In Sect.~\ref{sec:res}, we have observed a similar effect, where part of the tas signal was removed by conditioning on sst. However, it is important to note that this analysis cannot determine whether sst actually mediates the relationship between climate change and central European summer temperatures. This distinction will be discussed in Sect.~\ref{sub:corvscaus} below.

\subsection{Conditions act as a pure moderator}
While a mediator is part of the physical mechanism through which climate change modifies $X$, a moderator modulates the strength of the effect (Fig.~\ref{fig:modmed_diag}\,c). A simple example is when the climate change signal exhibits a seasonal cycle, in which case the season acts as a moderator of climate change. In our model (Eq.~(\ref{eq:modmedmodel})), this is represented by an interaction term between $S_i$ and $M$, i.e., $\beta_2\neq 0$.
In the simplest case, shown in Fig.~\ref{fig:modmed_diag}\,(c), all other causal connections are absent, i.e., $\Delta m=0$, $\beta_1=0$. The conditional and unconditional expectation changes are then related as 
\begin{align}
    E[X_1|m]-E[X_0|m] = (E[X_1]-E[X_0]) + \beta_2m\label{eq:deltaXmod}.
\end{align}
In addition, the moderator introduces a scenario dependence to the unconditional variance
\begin{align}
    \textit{Var}[X_1] =  \textit{Var}[X_0] + \beta^2\textit{Var}[\eta]\,,
\end{align}
which is removed, when we condition on $M$. These two effects are shown in the top row of Fig.~\ref{fig:modmed_pdfs}. Here, conditioning on a moderator increases the separation between distributions and affects the variances differently under the two scenarios, because $M$ contributes more variance under $S_1$.

While changing variances between scenarios were theoretically possible in our experiments (through the between model covariance, see Sect.~\ref{sub:mix}), they did not occur in Sect.~\ref{sec:res}. This suggests that possible moderation effects were neglected in our analysis. They might be represented in a more sophisticated statistical model, where the natural variability covariance matrix is permitted to differ between scenarios. 

\subsection{Correlation vs Causation}\label{sub:corvscaus}
A statistical model like Eq.~(\ref{eq:modmedmodel}) does not encode any information on the actual causal relations between the processes it represents. The parameter regimes we discussed above are consistent with the three causal pictures from Fig.~\ref{fig:modmed_diag}, but they would be equally consistent with different, possibly more complex situations, where for example $M$ and $X$ have other common drivers, or some of the directions of causality are reversed. Based on our results in Sect.~\ref{sec:res} alone, we cannot determine whether there is a physical pathway linking NH spring sst to NH summer temperatures, or whether the link is purely statistical. 

In a strict sense, the true causal structure can only be determined by an experiment in which the value of the conditions is prescribed. In the notation of causal theory proposed by \cite{pearl2009causality}, such experimentally constrained  probabilities are denoted as $\prob(X|\textit{do}(m))$, which is  generally not equal to $\prob(X|m)$. While a controlled experiment is impossible with the real climate system, it can be realised in silico using weather or climate model simulations.
For the effect of the scenario forcing $S$ itself, this was indeed realised by the historical and hist-nat simulations we used. The conditions under which these experiments allow for a causal interpretation are clarified in \cite{hannart2016}. 
In principle, the analogous experiment can be run for the effect of the weather or seasonal conditions $M$, for example by fixing the large-scale flow through data assimilation. This is the principle of the storyline school of climate change attribution \citep{shepherd2019storyline}. While an in-depth discussion of the different styles of conditional attribution is beyond the scope of this study, we can already see that this experimental style of conditioning can easily have effects that differ drastically from purely statistical conditioning: fixing a condition with no causal link from $M$ to $X$ will have no effect in a controlled storyline experiment. Conversely, it may heavily influence the outcome of our conditional statistics. This is particularly relevant in light of the third popular style of conditional attribution --flow analogues-- which also relies on statistical rather than experimental conditions. 

\section{Concluding remarks}\label{sec:outro}
Many recent attribution studies have begun to incorporate several complementary methodologies with varying kinds of conditioning. In this study, we have asked for the systematic effects of including such conditions. The NH summers of 2006 and 2007 serve as a striking example case: when we control for the NH summer blocking and, to a lesser degree, the conditions of the preceding NH spring sst, the probability ratio for tas in 2007 increases by orders of magnitude. In contrast, the summer of 2006 was not unusually warm once we account for the extremely favourable blocking circulation, leading to a conditional likelihood ratio close to 1. This illustrates that conditional and unconditional attribution approaches can differ systematically, and even lead to seemingly contradictory conclusions. 

Specifically, we have seen two main effects of conditioning. Firstly, removing part of the variability through conditioning leads to a better distinction between scenarios, resulting in more confident attribution statements. This reflects one of the original motivations for the storyline approach \citep{trenberth2011attribution}, namely the ability to analyse very rare and thus highly uncertain phenomena. In our example, the blocking index of \cite{sousa2021new} has proven to be a very strong predictor of NH summer temperatures in central Europe. By conditioning on this index, we explain around half of the natural variability and thereby greatly increase the separation between the two temperature distributions.
Secondly, the conditions shift our expectations of what is ``normal'' under each scenario. In our example, the exceptional blocking in 2006 caused the temperature distributions in both climates to shift towards higher temperatures. Consequently, the observed anomaly was actually more typical of the counterfactual climate, rather than being attributed to climate change.

As discussed in Sect.~\ref{sec:diss}, both of these effects can occur when our conditions act like a pure covariate, i.e., independently of climate change. The situation is significantly more complex when we allow for conditions that are involved in the causal chain, for example acting as moderators or mediators of climate change: conditioning on a mediator will ``regress'' out part of the mean signal. Conversely, a moderator condition may contribute additional variance under the factual scenario, which is removed when the condition is fixed. 

Although our conditional attribution experiments were based on low-dimensional normal distributions, many of the results can be applied to the conditional attribution frameworks that are commonly used in practice. In the case of flow analogues, for example, the condition variable describes how close the observed dynamical state of the atmosphere is to the target state. Depending on the presence of dynamical trends, this could be a moderator or mediator. 
As discussed in Sect.~\ref{sub:corvscaus}, the situation is fundamentally different for so-called dynamical storylines \citep{shepherd2019storyline}, where the conditions are experimentally enforced. We have argued that such experiments could be used to test whether the link between event and conditions is in fact causal or purely statistical. In the presence of dynamical trends, we should therefore not be surprised if analogues and storylines produce different results.

In this study, we considered climate change in the expectation values of normal distributions only. While this assumption is reasonable for seasonal averages and large geographic regions, it will break down when we move to more localised and short-lived events, such as individual storms or extreme rainfall events. Although the general ``Bayes theorem of probability ratios'' (Eq.~(\ref{eq:PRbayes})) still holds, the conditioning may have further consequences beyond those considered here. One example is the moderator effect, which would manifest as changes in the covariance structure under climate change. In the real world, the effective causal structure may be more complex with variables acting as both moderators and mediators, for example as moderators on shorter timescales and as mediators on longer timescales. Evidence for such effects could potentially be collected by applying models like Eq.~(\ref{eq:modmedmodel}) to data in a moderation and mediation analysis  \citep{igartua_mediation_2021}. Carefully controlled experiments similar to the storyline approach may then help in distinguishing between pure correlations an genuine causal pathways.


\appendix
\section{List of symbols}\label{app:symbols}
\begin{table}[!ht]
    \centering
    \caption{Selected symbols used throughout sections \ref{sec:theory} and \ref{sec:methods}. Estimators of some of these variables are denoted by at hat.}
    \begin{tabular}{ll}
         $\PR$ & probability ratio\\
         $\LR$ & likelihood ratio (ratio of probability densities)\\
         $S_1$, $S_0$& scenario with and without climate change\\
         $\RV{X},\RV{Y}$ & climate state vector in observations and models\\
         $\RV{\mu}_{t,i}$& mean climate state time $t$ under scenario $S_i$ \\
         $\RV{V}$& natural variability \\
         $\RV{\epsilon}$ & random observation error\\
         $\RV{b}_{t,i}$& climate model bias at time $t$ under scenario $S_i$\\
         $\mat{\Sigma}_{V,i}$ & covariance matrix of natural variability under $S_i$\\
         $\mat{\Sigma}_{j}$ & covariance matrix of natural variability for climate model $j$\\
         $\mat{\Sigma}_{b,t,i}$ & cov. matrix of model biases at time $t$ under $S_i$\\
         $\mat{\Sigma}_{\hat{\mu},t,i}$  & error cov. matrix of the estimated mean state at time $t$ under $S_i$ \\
         $\mat{\Sigma}_{\epsilon}$ & observation error cov. matrix\\
    \end{tabular}
    \label{tab:symbols}
\end{table}

\section{Blocking definition}\label{app:block}

We want to summarise the presence of atmospheric blocking by a single index which is strongly correlated with surface temperatures, can be localised over the region of interest and does not automatically contain a thermodynamic trend (for example due to the expansion of the atmosphere). These criteria are met by the relatively recent approach of \cite{sousa2021new} based on daily 500 hPa geopotential height fields. Here, the effect of thermal expansion is removed by considering only spatial gradients and anomalies with respect to a short time window. A key novelty of this blocking definition is the identification of subtropical ridges in addition to classic Rex- or Omega-type blocks. Grid points are considered instantaneously blocked if either (a) the geostrophic flow in 500 hPa is reversed (Rex or Omega, similar to \cite{davini2012bidimensional}) or (b) a ridge with geopotential height above the hemispheric mean in the preceding 15 days extends north of the edge of the subtropical belt. Contiguous blocking objects are identified in the daily fields, filtered for sizes $\geq 500.000$ km$^2$ and then tracked over time to ensure a lifetime of at least four days. For the details of the algorithm, we refer to \cite{sousa2021new}. The resulting daily binary blocking fields are then averaged over the study region and each summer to obtain a yearly timeseries of blocking activity. 

\section{Gaussian LR in case of constant variances} \label{app:toy}

In general, the logarithm of the normal pdf under scenario $S_i$ can be written as
\begin{align}
    \log f_i(x) = -\frac{(x-\mu_i)^2}{2\sigma_i^2}-\frac{1}{2}\log(2\pi) - \log(\sigma_i)\,.
\end{align}
Assuming equal variances under the two scenarios, we find for log-likelihood ratio
\begin{align}
    \log \LR &= \log f_1(x)-\log f_0(x)\\
    &=\frac{1}{2\sigma^2}\cdot\left( (x-\mu_0)^2 - (x-\mu_1)^2 \right)\\
    &=\frac{1}{2\sigma^2}\cdot\left( x^2-2\mu_0x+\mu_0^2-x^2+2\mu_1x-\mu_1^2 \right)\\
    &=\frac{1}{2\sigma^2}\cdot\left( 2x\cdot(\mu_1-\mu_0)+\underbrace{\mu_0^2-\mu_1^2}_{=(\mu_0-\mu_1)(\mu_0+\mu_1)} \right)\\
    &=\frac{1}{\sigma^2}\cdot (\mu_1-\mu_0)\cdot\left(x-\frac{\mu_0+\mu_1}{2}\right) \label{eq:normalPR}
\end{align}
The outcome of our attribution study is thus determined by the variance of the distribution, the climate change signal in the mean and the anomaly of the event with respect to the average mean state.

Noting that the conditional distribution of $X_i|M_i=m$ is normal, we obtain Eq.~(\ref{eq:logPRcov}) for the pure covariate case by inserting the expressions for the conditional means and variances (Eqs.~(\ref{eq:E|M}) and (\ref{eq:V|M})) into Eq.~(\ref{eq:normalPR}).




\section*{Acknowledgements}
This research was funded within the BMBF project ClimXtreme phase 1 under grant number 01LP1902A and  ClimXtreme phase 2 under grant number 01LP2323A. This work used resources of the Deutsches Klimarechenzentrum (DKRZ) granted by its Scientific Steering Committee (WLA) under project ID 1152. We are especially grateful to Etor Lucio Eceiza at DKRZ for his invaluable technical support. We furthermore thank Pedro Sousa for advice on his blocking definition and Ieda Pscheidt-Willems for her contributions in the early phases of this work.


\bibliography{sources.bib}

\end{document}